\documentclass[twocolumn,superscriptaddress, floatfix, showpacs, aps, prb, 10pt]{revtex4-2}
\usepackage{graphicx}
\usepackage{amsmath}
\usepackage[colorlinks=true, citecolor=blue, urlcolor=blue]{hyperref}
\usepackage{amssymb}
\usepackage{bm}
\usepackage{color}
\usepackage{array}
\usepackage{booktabs}
\usepackage{multirow}
\usepackage{braket}

\begin{document}

\title{Electronic structure of the surface superconducting Weyl semimetal PtBi$_2$}

\author{Riccardo Vocaturo}
\affiliation{Leibniz Institute for Solid State and Materials Research (IFW) Dresden, Helmholtzstrasse 20, 01069 Dresden, Germany}

\author{Klaus Koepernik}
\affiliation{Leibniz Institute for Solid State and Materials Research (IFW) Dresden, Helmholtzstrasse 20, 01069 Dresden, Germany}

\author{Jorge I. Facio}
\affiliation{Centro Atomico Bariloche, Instituto de Nanociencia y Nanotecnologia (CNEA-CONICET) and Instituto Balseiro, Av. Bustillo, 9500, Argentina}

\author{Carsten~Timm}
\affiliation{Institute of Theoretical Physics, Technische Universit\"{a}t Dresden, 01062 Dresden, Germany}
\affiliation{W{\"u}rzburg-Dresden Cluster of Excellence ct.qmat, Helmholtzstrasse 20, 01069 Dresden, Germany}

\author{Ion Cosma Fulga}
\affiliation{Leibniz Institute for Solid State and Materials Research (IFW) Dresden, Helmholtzstrasse 20, 01069 Dresden, Germany}
\affiliation{W{\"u}rzburg-Dresden Cluster of Excellence ct.qmat, Helmholtzstrasse 20, 01069 Dresden, Germany}

\author{Oleg Janson}
\affiliation{Leibniz Institute for Solid State and Materials Research (IFW) Dresden, Helmholtzstrasse 20, 01069 Dresden, Germany}

\author{Jeroen van den Brink}
\affiliation{Leibniz Institute for Solid State and Materials Research (IFW) Dresden, Helmholtzstrasse 20, 01069 Dresden, Germany}
\affiliation{Institute of Theoretical Physics, Technische Universit\"{a}t Dresden, 01062 Dresden, Germany}
\affiliation{W{\"u}rzburg-Dresden Cluster of Excellence ct.qmat, Helmholtzstrasse 20, 01069 Dresden, Germany}

\begin{abstract}
Trigonal PtBi$_2$ is a layered semimetal without inversion symmetry, featuring 12 Weyl points in the vicinity of the Fermi energy. 
Its topological Fermi arcs were recently shown to superconduct at low temperatures where bulk superconductivity is absent. 
Here, we perform first-principles calculations to investigate in detail the bulk and surface electronic structure of PtBi$_2$, and obtain the spin texture as well as the momentum-dependent localization of the arcs. 
Motivated by the experimentally observed recovery of inversion symmetry under pressure or upon doping, we interpolate between the two structures and determine the energy and momentum dependence of the Weyl nodes. 
For deeper insights into the surface superconductivity of PtBi$_2$, we construct a symmetry-adapted effective four-band model that accurately reproduces the Weyl points of PtBi$_2$. 
We supplement this model with an analysis of the symmetry-allowed pairings between the Fermi arcs, which naturally mix spin-singlet and spin-triplet channels. 
Moreover, the presence of surface-only superconductivity facilitates an intrinsic superconductor-semimetal-superconductor Josephson junction, with the semimetallic phase sandwiched between the two superconducting surfaces. 
For a phase difference of $\pi$, zero-energy Andreev bound states develop between the two terminations.
\end{abstract}

\maketitle

\section{Introduction}
\label{sec:introduction}

It was recently established experimentally that trigonal PtBi$_2$ is a layered material that combines topological features of its electronic structure, in particular, the presence of Weyl nodes, with the occurrence of superconductivity \cite{Veyrat2023, Kuibarov2023}. 
The Weyl nodes in this non-magnetic system arise due to the absence of inversion symmetry and they are of type I \cite{Veyrat2023}. 
Angle-resolved photoemission spectroscopy (ARPES) has clearly identified the topological surface states related to the Weyl nodes, i.e. the Fermi arcs, and it has also shown that these Fermi arcs become superconducting at a temperature for which bulk superconductivity is absent \cite{Kuibarov2023}. 
At one type of surface termination, the superconducting critical temperature is found to be $T_c = 14\,\mathrm{K}$; for the other surface termination, it is $8\,\mathrm{K}$.
Scanning tunneling spectroscopy experiments confirm the presence of surface superconductivity from the observation of superconducting gaps \cite{Schimmel2023}. 
However, the measured gap size depends on the surface location and it can even reach values of $20\,\mathrm{meV}$.
Within the BCS theory, this suggests a $T_c$ possibly one order of magnitude higher than the ones measured by ARPES. 
In this context, it is remarkable that in thin exfoliated flakes of PtBi$_2$ a Berezinskii-Kosterlitz-Thouless-type superconducting transition has been observed at temperatures below $0.4\,\mathrm{K}$ \cite{Veyrat2023}. 
An inhomogeneous (not bulk) superconducting phase with $T_c=1.1\,\mathrm{K}$ has also been recently reported \cite{Zabala_2024}. 
Moreover, electron transport through a small metallic constriction at the surface, i.e. a point contact, evidences superconductivity with a critical temperature in the range $2.5-7.5\,\mathrm{K}$, depending on the metal used for the point contact~\cite{Bashlakov2022}.

Different pieces of experimental evidence have the common denominator that the surface of PtBi$_2$ is superconducting in a temperature regime at which the bulk remains metallic.
At the same time, a consistent and detailed interpretation of this set of observations is still lacking.
Theoretically, if the surface carries a higher density of states compared to the bulk, it may also have a higher mean-field superconducting transition temperature, in spite of its lower dimensionality \cite{Nomani2023}.

The experimentally observed Fermi arcs in ARPES compare well with the findings of \textit{ab initio} calculations of the electronic band-structure, as does the bulk spectral function \cite{Kuibarov2023}.
Introducing superconductivity in the computational framework via the Bogoliubov--de Gennes formalism, using a nonzero pairing potential only close to the surface, the observed superconducting gap formation in the Fermi arcs can be phenomenologically described.
The surface-only superconductivity in the PtBi$_2$ Fermi arcs is of particular interest as it may permit to manipulate topological and superconducting phases in a single material. 
For instance, varying the flake thickness of a single crystal can lead to an intrinsic tunable Josephson junction with a topological Weyl semimetal forming the weak link.

At present, the open questions on PtBi$_2$ not only concern the microscopic origin, homogeneity, symmetry, and mechanism of the surface superconductivity but also its relation to the topological nature of the Fermi arcs, and, by extension, to the role of bulk inversion-symmetry breaking.  
Indeed, when inversion symmetry is restored, for instance, through doping with a small concentration of Te or Se or by applying pressure, a robust \textit{bulk} superconducting state emerges with $T_c \sim 2\,\mathrm{K}$ \cite{Takaki2022, Wang2021}, in the absence of topological Fermi arcs. 
Moreover, the topological electronic states appear to play an important role for superconductivity also in other materials. 
In point-contact spectroscopy, superconductivity can be induced at the surface of a wide range of non-superconducting bulk semimetals, including the Weyl semimetal TaAs \cite{Aggarwal2017, Wang20177, Hou2020}, the three-dimensional Dirac semimetal 
Cd$_3$As$_2$ \cite{Wang2015, Aggarwal2015}, the line-nodal semimetal ZrSiS \cite{Aggarwal_2019}, the topological crystalline insulator Pb$_{0.6}$Sn$_{0.4}$Te \cite{Das2016}, and the triple-point topological semimetal tungsten carbide WC \cite{Hou2019}. 
This raises the question whether there is an intrinsic surface superconducting instability to these topological electronic structures.

Here, we focus on the electronic structure of PtBi$_2$ from a theoretical perspective, considering both the normal and (surface) superconducting state. 
We consider the spin textures of the Weyl nodes and Fermi arcs, as well as their evolution under pressure and doping. 
The trajectory of the Weyl nodes (both in energy and momentum space) as we turn from the inversion-symmetry broken to the centrosymmetric crystal structure is obtained. 
An effective model with the minimum number of four bands is proposed to describe the metallic phase, aiming at reproducing the topological features that stem from the presence of Weyl points close to the Fermi level.
We also discuss the superconducting pairing state at a single surface of PtBi$_2$, restricting ourselves to superconducting states that do not break crystalline symmetries. 
We identify the three states that span the space of the possible superconducting order parameters: $s$-wave spin-singlet pairing, $f$-wave spin-triplet pairing of Ising type, and $p$-wave spin-triplet pairing of Rashba type.
Finally, we consider the possibility of an intrinsic superconductor-semimetal-superconductor Josephson junction, where the semimetallic phase is sandwiched between the two superconducting surfaces. 
We find in-gap Andreev bound states if a phase difference develops between the two terminations.

This article is structured as follows. After briefly introducing Weyl semimetals (Sec.~\ref{sec:intro_Weyl}), the crystal structure of PtBi$_2$ is discussed in Sec.~\ref{sec:structure}. 
Section \ref{sec:dft} is dedicated to first-principles electronic-structure calculations. 
In Sec.~\ref{sec:model}, we present our effective model and highlight its main features, after which we analyse the superconducting pairing channels in Sec.~\ref{sec:pairing}. 
In Sec.~\ref{sec:junction}, we investigate the Andreev zero modes in an intrinsic $\pi$ junction between the two superconducting surfaces. 
Finally, we summarize our findings in Sec.~\ref{sec:conclusion}.

\section{Weyl semimetals and Fermi arcs}
\label{sec:intro_Weyl}

To provide a general context, we conceptually introduce Weyl semimetals and their Fermi-arc surface states. 
This topological state of matter can be understood from slicing its three-dimensional Brillouin zone (BZ) into planes, thereby fixing one of the three components of the momentum. 
The resulting two-dimensional (2D) band structure will generally display a set of band gaps. 
For each gap, one can determine whether the corresponding 2D insulating state is topologically trivial or not by calculating the corresponding Chern number. 
Then, the gap characterizing the two-dimensional slice of BZ can be followed adiabatically along the third direction. 
Should the gap of the slice change character from trivial to topological along this path, or vice versa, there must be a nodal point in momentum space where the band gap closes. 
Around these nodes, which in a periodic system are forced to appear as pairs with opposite chirality \cite{Nielsen1981a, Nielsen1981b}, the dispersion is linear. 
They are referred as Weyl points. 
Together, the 1D topological edge states of each 2D topologically insulating slice build a 2D plane of surface states with energy depending on the two surface momenta. 
At fixed energy, these states appear as arcs in momentum space, connecting the 2D projection of the bulk Weyl points. If they cross the Fermi energy, they are referred as Fermi arcs.

A prerequisite for Weyl nodes is to break either inversion symmetry, or time-reversal symmetry, or both. 
The nodes are robust against perturbations: Weyl nodes can only annihilate if an external perturbation makes pairs of nodes of opposite chirality meet at the same point in the Brillouin zone so that they can gap each other out. 
Conceptually the simplest Weyl semimetal would be realized by a system with Weyl nodes close to the Fermi level so that each node has a spherical Fermi-surface around it, where the chirality of the node corresponds to the Chern number of its Fermi-surface ball. 
However, in real materials the semimetallic nature enforced by these nodes may coexist with other bands crossing the Fermi level, complicating the situation. 

Beyond this, superconductivity many lead to further complications, even in the presence of time-reversal symmetry. 
Generally speaking, it is allowed to have pairings of different symmetries between electrons on different Fermi pockets, with the same or different Chern number. 
In such cases, topological superconductivity can already appear in the $s-$wave pairing channel \cite{Qi2010, Hosur2014, Li2018}. 
Moreover the topological edges states themselves, the Fermi-arcs, may exhibit an independent superconducting instability \cite{Nomani2023}.

\section{Crystal structure}
\label{sec:structure}

\begin{figure}[tb]
    \centering
    \includegraphics[width=0.49\textwidth]{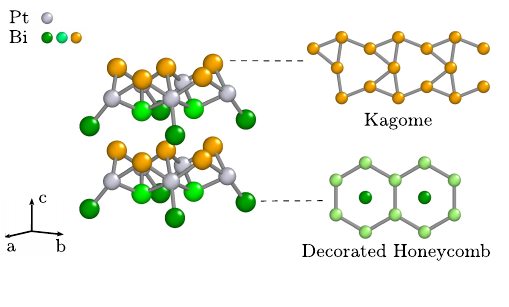}
    \caption{Crystal structure of PtBi$_2$ described by the space group $P31m$ (157). 
    Colors (green, light green and orange) denote three inequivalent Wyckoff positions of Bi. Each Bi-Pt-Bi layer has two inequivalent terminations: a flat monolayer with the kagome structure (top right) and a
corrugated monolayer with a decorated honeycomb structure (bottom right).}
    \label{fig:structure}
\end{figure}

The crystal structure of pristine PtBi$_2$ is described by the noncentrosymmetric space group $P31m$ (157) \cite{ptbi2:kaiser14, Shipunov2020} which features three mirror planes (along $a$, $b$, and the short diagonal) whose intersection gives rise to the $C_3$ axis. 
This structure can be obtained from the more symmetric CdI$_2$-type phase, upon subsequent splitting of the (Bi) $2d$ Wyckoff position, inducing a 60$^{\circ}$ rotation of the mirror planes and polarity along the stacking direction [001] \cite{ptbi2:kaiser14}. 
Therefore, PtBi$_2$ is characterized by Bi-Pt-Bi layers, with markedly different Bi terminations, see Fig.~\ref{fig:structure}: while one of the Bi monolayers is nearly flat and its shortest Bi-Bi connections form a kagome lattice, the other monolayer is strongly corrugated and can be described as a decorated honeycomb lattice. 
The inequivalence of the Bi monolayers plays a crucial role for different characteristics of the surface states, as we will show later.
Finally, we note that also the Pt monolayers feature uneven
interatomic separations: three neighboring Pt atoms form a regular triangle with metallic bonds~\cite{ptbi2:kaiser14}; in turn such triangles are arranged into a triangular lattice.

Pressure or chemical substitution of a small fraction of Bi by Se or Te induces a transition to a centrosymmetric trigonal structure (space group $P\bar3m1$), which we discuss in more detail in Sec.~\ref{sec:centro_struc}.

\section{Electronic structure from first principles}
\label{sec:dft}

To study the electronic structure of PtBi$_2$, we performed full-relativistic density functional theory (DFT) calculations using the full-potential local-orbital code FPLO~\cite{Koepernik1999} version 22.01-63. 
For the exchange-correlation term, we used the generalized gradient approximation (GGA)~\cite{Perdew1997}. 
The Brillouin zone was sampled by a $12\times12\times12$ mesh of $k$-points. 
(We will comment on the \textit{k}-mesh influence later in this Section, in the discussion of Weyl points.) 
The experimental crystal structure from Ref.~\onlinecite{Shipunov2020} was used as the structural input. 
For slab calculations, used the dedicated FPLO module~\cite{koepernik23} to construct a Wannier Hamiltonian by projecting Kohn-Sham states onto a basis of 72 spin-orbitals that comprises Pt $6s$, $5d$, and Bi
$6p$ states. 
The Wannier projection reproduces the full band structure from $-7$ to $5\,\mathrm{eV}$.
When needed, relaxation of internal atomic coordinates without space-group-symmetry constraints was performed. 
For the noncentrosymmetric structure this translates to relaxing the \textit{z}-coordinates of all Wyckoff positions, as well as the \textit{x}-coordinates of Wyckoff position 6d and 1a.
For the inversion symmetric structure only Bi's \textit{z}-coordinates was allowed to relax.
Further discussions are found in the corresponding sections (\ref{sec:centro_struc}).

\subsection{Bulk band structure}

\begin{figure}[tb]
    \centering
    \includegraphics[width=0.5\textwidth]{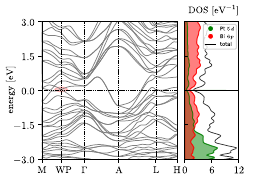}
    \caption{Full-relativistic GGA band structure and density of states (DOS) of bulk PtBi$_2$. 
    The momentum of the highlighted Weyl point (WP) deviates slightly from $\Gamma$--M path. 
    For the DOS, a Gaussian smearing of $0.05\,\mathrm{eV}$ was used.}
    \label{fig:fatbanddos}
\end{figure}

The GGA band structure and the corresponding density of states are shown in Fig.~\ref{fig:fatbanddos}. 
In the vicinity of the Fermi level, Pt $5d$ and Bi $6p$ contributions are dominant. 
The former contribute primarily to the occupied part of the spectrum, and give rise to flatter bands. 
By contrast, Bi $6p$ bands are more dispersing and span a much wider range of energies.

States around the Fermi level are mostly Bi $6p$, with a minor admixture of Pt $5p$. 
The DOS, see Fig.~\ref{fig:fatbanddos} (right) features a valley around the Fermi level -- visible also in orbital-resolved DOS contributions -- signalling the semimetallic behavior. 
Interestingly, we observe a significant dispersion along $\Gamma$--A, indicating that the electronic structure of this layered material is three-dimensional. 
Moreover, the shape of the band crossing at the Fermi level is typical for chains with second-neighbor couplings, suggesting the presence of interlayer couplings beyond nearest neighbors.

The Fermi surface of PtBi$_2$ is strongly three-dimensional, and consists of several distinct sheets mostly located close to the A--L--H plane. 
This results agrees with ARPES~\cite{Kuibarov2023, Jiang2020} and de Haas--van Alphen experiments \cite{Veyrat2023, Gao2018} as well as earlier band structure calculations~\cite{Kuibarov2023, Gao2018, Jiang2020}. 
Nevertheless, specifics of the Fermi surface are sensitive to computational details, and despite the overall good agreement, we observe discrepancies between Refs.~\onlinecite{Jiang2020} and \onlinecite{Gao2018}. 
Our calculations conforms to the ones presented in Ref.~\onlinecite{Veyrat2023} (in supplementary). 
We identify the same sheets as in Ref.~\onlinecite{Jiang2020}, however, the Fermi surface referred to as B.III (in Ref.~\onlinecite{Jiang2020}) is found to be disconnected around the $k_z=0$ plane, as reported in Ref.~\onlinecite{Gao2018}. 
This is due to the very small gap in the band structure between $\Gamma$--M, which is sensitive to the exact position of the Fermi level.

\begin{figure}[tb]
    \centering
    \includegraphics[width=0.49\textwidth]{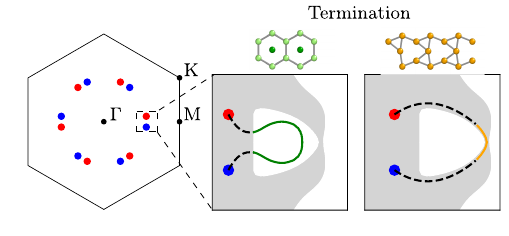}
    \caption{Projection of 12 symmetry-connected Weyl points onto the $k_z=0$ plane of the Brillouin zone. 
    Red and blue colors refer to positive and negative chirality. 
    The blowups show a sketch of the surface BZ, where the projections of a pair of WPs are connected by a Fermi arc. 
    Following Fig.~\ref{fig:structure}, the green (orange) color corresponds to decorated honeycomb (kagome) Bi surfaces; the shaded grey area denotes bulk states.
Dashed lines are a guide to the eye (Fermi arcs are traceable only in regions of 2D BZ that are free of bulk states).}
    \label{fig:wppos}
\end{figure}

The lack of inversion symmetry in PtBi$_2$ generally allows for linear band crossings -- Weyl points. 
Notably, pairs of Weyl points are located $48\,\mathrm{meV}$ above the Fermi energy, with Fermi arcs crossing the Fermi level \cite{Kuibarov2023}. 
One such point is located at $k\!=\!(0.324,0.041,-0.153)$
in units of $(2\pi/a,2\pi/b,2\pi/c)$. 
This and further eleven symmetry-related Weyl points are schematically depicted in Fig.~\ref{fig:wppos}. 
The exact values of momenta and energies are indeed very sensitive to small variations of the crystal structure, in particular to the interlayer distance between the Bi planes. 
Moreover, we studied the energy dependence of the Weyl points on the \textit{k}-mesh up to a $21 \times 21 \times 19$ grid ($\approx 8400$ points), for which the energy of the Weyl points reaches $43\,\mathrm{meV}$, i.\ e.\ it drops by $5\,\mathrm{meV}$ compared to reference ($12\times12\times12$) calculation. 
The dispersion in all three momentum directions is linear, and since only a very slight tilt of the cone is observed, PtBi$_2$ is classified as a type-I Weyl semimetal. 
We note that, in our investigation, we focused on the crossings that are closest to the Fermi level, but the same bands yield other intersections at higher energies. 
For example, other six pairs are found at $150\,\mathrm{meV}$; the coordinates of one of such points being $(0.3425,0.1044,0.2606)$ in units of $(2\pi/a,2\pi/b,2\pi/c)$ (calculated with GGA on the same $12 \times 12 \times 12$ mesh).

\subsection{Weyl points}

\begin{figure}[tb]
    \centering
    \includegraphics[width=0.5\textwidth]{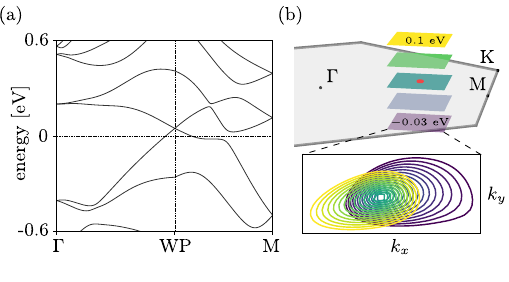}
    \caption{(a) Band structure in the vicinity of the Weyl point. 
    (b) Sketch of $k_z=k_z^{\text{WP}}$ planes at different energies at which constant energy cuts were collected (not in scale). 
    The blowup (bottom) shows constant energy cuts for these and other intermediate energy values between $-0.03$ and $0.1\,\mathrm{eV}$. 
    The weak tilting of the cone agrees with Ref.~\onlinecite{Veyrat2023}. 
    }
    \label{fig:wpzoom}
\end{figure}

Weyl points are situated near the $\Gamma$--M line. One of the respective bands with pronounced Pt $5d$ character is remarkably flat in the corresponding part of the BZ (Fig.~\ref{fig:wpzoom}). 
Indeed, Pt $5d$ contributions to these cones are more than 30\% larger than the momentum-averaged value from the DOS calculations shown in Fig.~\ref{fig:fatbanddos}. 
This indicates that hybridization between atomic orbitals
is strongly momentum-dependent.

Inspecting the dominant $6p$ Bi contribution, we find that Bi atoms belonging to the kagome plane take about half of the relative $6p$ weight, while the other Bi atoms account for the other half almost equally.
As a result, it is challenging to single out a specific set of orbitals giving rise to the Weyl crossing. 
Nevertheless, we observe a prevalence of in-plane orbitals $p_x$ and $p_y$ over $p_z$ orbitals.

\begin{figure}[tb]
    \centering
    \includegraphics[trim={0cm 0cm 0 0cm}, clip, width=0.5\textwidth]{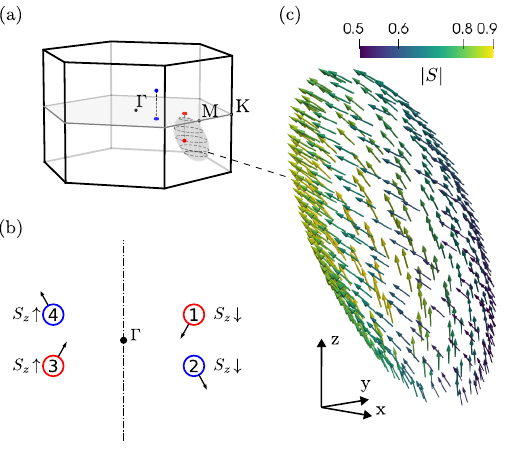}
    \caption{(a) Schematic representation of the constant energy ($0.1\,\mathrm{meV}$ below the energy of the Weyl point) surface in \textit{k}-space (not to scale) around the Weyl point. 
    (b) Schematic illustration of spin projections of symmetry-equivalent WPs (not to scale). 
    The arrows connected to the colored circles represent the in-plane spin components $S_x$ and $S_y$. 
    The dash-dotted line represents a mirror plane. See text for details. 
    (c) Semi-classical representation of the spin over the surface shown in (a); $(x,y,z)$ refer to spin directions. 
    The colors represent the magnitude of $S$ in units of $\mu_B$, which is not conserved in the presence of spin-orbit coupling.}
    \label{fig:spin_wp}
\end{figure}

We also examined the spin texture of the bands forming the cones. 
It was found to smoothly evolve around the cone without any distinct correlation to the singularity in the Berry curvature. We show in Fig.~\ref{fig:spin_wp}(c) the spin texture of the constant energy surface located $0.1\,\mathrm{meV}$ below the energy of the Weyl point. 
Due to the slight tilt of the cone, this surface is an ellipsoid, and not a regular sphere. 
For completeness, we repeated our analysis for a slightly different energy ($0.3\,\mathrm{meV}$) and found no qualitative difference in the spin texture.

In Fig.~\ref{fig:spin_wp}(b), we show how the spins of the Weyl points with opposite chirality are connected by symmetries, plotting the effect of these symmetries on the average spin vector $\left\langle {\mathbf S} \right\rangle$. 
WP \#1 features the spin structure shown in Fig.~\ref{fig:spin_wp}(c). 
Time-reversal symmetry relates WP \#1 to WP \#3 at $-\mathbf k$, flipping all three spin components but leaving chirality unchanged (red color). 
Subsequently, through the mirror symmetry along the $y$-axis, we reach WP \#2 with opposite chirality. 
As mirrors preserve only the component perpendicular to the plane of reflection, both $S_y$ and $S_z$ acquire an extra negative sign besides the one induced by time reversal. 
Consequently, the neighboring pair of Weyl points
share the same $S_y$ and $S_z$, but opposite $S_x$. 
In this case, WP \#1 has $\left\langle {\mathbf S} \right\rangle \approx (-0.2,-0.36,0.5) \mu_B$.

Lastly, we note that the spin structure of Weyl points
can play an important role for the dielectric response~\cite{Timm2023}: the long-distance behavior of Friedel oscillations is strongly influenced by the
relative spin structure of a pair of Weyl points, giving rise to singularities in the derivatives of the dielectric function. 

\subsection{Evolution of Weyl points under pressure and doping}
\label{sec:centro_struc}

\begin{figure}[tb]
    \centering
    \includegraphics[width=0.49\textwidth]{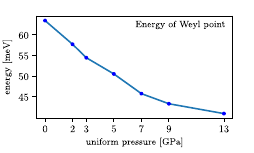}
    \caption{Energy of the Weyl point for the optimized structures as a function of pressure. 
    For calculations, a $k$-mesh of 6000 points has been used. 
    To convert volume into pressure we used the equation of state from Ref.~\onlinecite{Wang2021}. 
    At $13\,\mathrm{GPa}$, $V/V_0 = 0.92$.}
    \label{fig:wp_vs_e}
\end{figure}

The noncentrosymmetric $P31m$ structure of PtBi$_2$ is stable up to an external hydrostatic pressure of about 13 GPa, with a remarkably resilient $c/a$ ratio of about $0.94$~\cite{Wang2021}. 
The smooth evolution of lattice parameters allows us to investigate the momentum and energy of the Weyl point as a
function of pressure. 
To this end, we used the experimental lattice constants from Ref.~\onlinecite{Wang2021}, and optimized the internal coordinates with scalar relativistic calculations and a convergence threshold of $10^{-5}\,\mathrm{eV}/\mathrm{\AA}$. 
Structural optimizations are necessary, because the pressure dependence of the atomic positions is unknown. 
For consistency, the structure at zero pressure was also relaxed. 
Note that it gives rise to a shift of the Weyl point's energy to $65\,\mathrm{meV}$ (compared to $48\,\mathrm{meV}$ for the experimental structure).
With the optimized structure, a full relativistic calculation followed by a Wannier projection were performed to then search for the Weyl points. 
While the relaxation induces minute changes in the atomic coordinates, the exact position and energy of the Weyl points are sensitive to these variations, as well as to the choice of \textit{k}-mesh (as expected). 
The \textit{k}-mesh was scaled accordingly to the changes in the unit cell's volume to maintain the same number of points per unit volume. 
Convergence was reached at 6000 \textit{k}-points. 
We observe a sizable drop in energy (about $20\,\mathrm{meV}$) in the investigated pressure range, as shown in Fig.~\ref{fig:wp_vs_e}. 

To investigate whether the Weyl points could be brought even closer to the Fermi level, we estimated the additional charge needed to shift the chemical potential by integrating the DOS from the Fermi energy up to the energy of the Weyl point. 
To avoid regions potentially affected by structural instabilities, we restrict our comparison to the values at $0$ and $9\,\mathrm{GPa}$: $0.013$ and $0.0098\, e/\mathrm{atom}$, respectively. 
This yields an average doping of $0.01\, e/\mathrm{atom}$ (or $\approx 0.1$ additional unit of charge per unit cell) required to bring the Weyl point down to the Fermi level.  
This level of doping can be achieved by a small excess of Bi as well as doping by Se or Te. 
Interestingly, in Ref.~\onlinecite{Takaki2022}, the authors report a transition to a centrosymmetric trigonal structure (space group $P\bar3m1$) upon minimal Se or Te doping. 
The phase transition is found at $x=0.024$ for Te and $x=0.032$ for Se, where $x$ refers to the stoichiometry in the notation Pt(Bi$_{1-x}$Ch$_x$)$_2$, $\mathrm{Ch} = \mathrm{Se}, \mathrm{Te}$. 
Thus, the overall additional charge per atom is $(x \times 2)/ 3 \approx 0.016,0.021\, e/\mathrm{atom}$, which is very close to our estimated charge  necessary to bring the Weyl points down to the Fermi level. 

\begin{figure}[tb]
    \centering
    \includegraphics[width=0.46\textwidth]{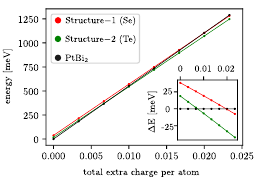}
    \caption{Total energy contribution induced by doping PtBi$_2$ (computed within GGA). 
    To account for small energy differences, we used on a fine mesh of 8000 \textit{k}-points. 
    In the inset, we plot the relative energy difference, using PtBi$_2$ as a reference.}
    \label{fig:dop}
\end{figure}

To further explore this scenario, we performed additional total energy calculations for the noncentrosymmetric and centrosymmetric PtBi$_2$ phases, using the experimental lattice parameters~\cite{Takaki2022} and relaxing the
internal coordinates. 
Given the small concentrations of Se/Te, we mimic the effect of doping by adding a homogeneous positively-charged background (jellium model). 
We note that in our approach the Se- and Te-doped structures differ only by their unit cell volume, following the different atomic radii of Se and Te. 
Results are shown in Fig.~\ref{fig:dop}. 
We found the centrosymmetric structure to become energetically favorable at $\sim\!0.007$ and $\sim\!0.018\, e/\mathrm{atoms}$ for Te and Se, respectively. 
This translates to $x \approx 0.01,0.027$, which is in good agreement with experiments ($x=0.024,0.032$)~\cite{Takaki2022}. 
Similarly, the transition is shifted to lower doping for the Te-doped structure featuring a larger unit cell volume.

\begin{figure}[tb]
    \centering
    \includegraphics[width=0.5\textwidth]{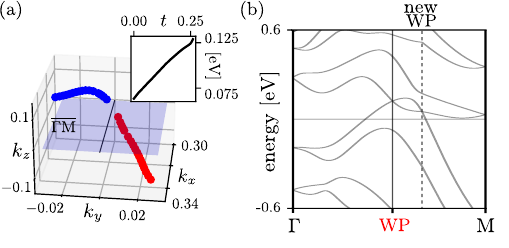}
    \caption{(a) Merging of a pair of Weyl points of opposite chirality along the interpolating path between the inversion broken ($P31m$) and centrosymmetric ($P\bar{3}m1$) structure. 
    The axes are given in units of $2\pi/a$. 
    The merging occurs on the high-symmetry $\overline{\Gamma \mbox{M}}$ line for $t\approx 0.26$. 
    The inset shows the energy dependence as a function of $t$. 
    (b) Band structure for $t=0.24$: right before the Weyl points annihilate, a new pair emerges in the same band.}
    \label{fig:path}
\end{figure}

Given the structural proximity of these phases, it is of interest to look at the evolution of the Weyl points along a continuous path connecting the two crystal structures. 
To this end, we took the experimental noncentrosymmetric and centrosymmetric (Se-doped) structures, and relaxed the internal atomic coordinates. 
These are the endpoints of the path. 
This additional relaxation step was necessary to minimize numerical noise in the energies and momenta of the Weyl points, as previously discussed. 
Then, we proceeded to sample the path connecting the two relaxed structures with a linear interpolation of the atomic coordinates and lattice constants. 
The interpolating parameter is referred as \textit{t}: for $t=0$ we have the inversion-broken structure, while at $t=1$ we find the centrosymmetric one. 
At each t, an intermediate structure is obtained for which we performed full-relativistic calculations and Wannier projection in order to locate the Weyl points. 
The results are summarized in Fig.~\ref{fig:path}.

We observe the WPs moving closer to the $\Gamma$M line, where they merge for $t \approx 0.26$. 
This behavior is expected since the pair's coordinates are related to each other by time reversal and mirror symmetry, enforcing the merging to happen on this line (at $k_z=0$). 
Around $t=0.24$, slightly before the annihilation process is completed, a new pair is formed, closing the inverted gap which we can see in Fig.~\ref{fig:wpzoom} in the vicinity of the M point. 
This new pair follows a more involved trajectory.

\subsection{Fermi arcs}

\begin{figure}[tb]
    \centering
    \includegraphics[width=0.48\textwidth]{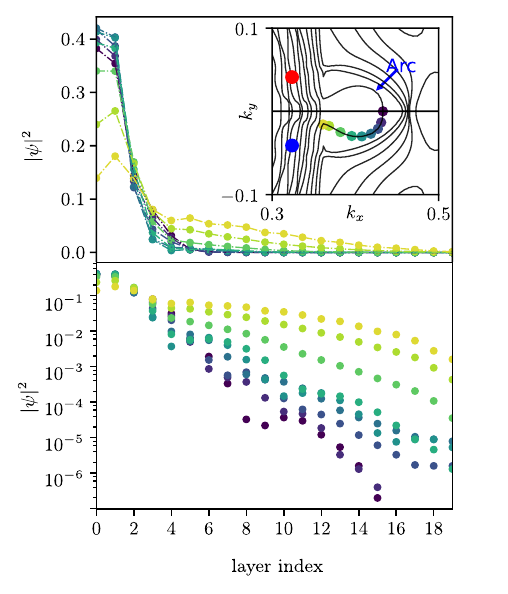}
    \caption{Localization of the Fermi arc on the decorated honeycomb surface of a finite slab comprising 20 PtBi$_2$ layers. 
    The probability density of the Fermi arc wavefunction is plotted as a function of layer index.
    Colors correspond to the positions of $k$ points in the Fermi surface shown in two-dimensional Brillouin zone in the inset. 
    Red and blue circles indicate the projections of Weyl points with opposite chirality. 
    Bottom: same data in a logarithmic scale.}
    \label{fig:loc}
\end{figure}

A hallmark of Weyl semimetals are Fermi arcs -- topologically protected surface states. 
These Fermi arcs were observed experimentally in ARPES, in excellent agreement with DFT-calculated energy distribution curves~\cite{Kuibarov2023}.
Here, we restrict ourselves to a schematic representation in Fig.~\ref{fig:wppos}. 
An interesting characteristic of the Fermi arcs in PtBi$_2$
is their sensitivity to the surface termination. 
While Fermi arcs on the kagome surface are strongly hybridized with adjacent bulk states, the decorated honeycomb surface exhibits distinct Fermi arcs isolated from the neighbouring bands (Fig.~\ref{fig:loc}, inset). 
Thus, we will focus primarily on the latter termination.

To study the localization of the states forming the Fermi arcs, we constructed a Wannier Hamiltonian on a finite slab comprising 20 PtBi$_2$ layers.
Subsequently, we grouped the atomic contributions from each layer and projected the eigenvectors onto these states at various $\mathbf k$-points along the arc. 
The calculated localization for the decorated honeycomb surface is shown in Fig.~\ref{fig:loc}. 
In large parts of the two-dimensional Brillouin zone the Fermi arcs are well-separated from other slab states, allowing us to study the penetration of the surface states
into the bulk as a function of $\mathbf k$. 
As expected, a distinct peak is observed within the first two to three layers for all the points investigated. 
The density rapidly drops for deeper layers, with the rate of decrease depending on $\mathbf k$: as we approach the projection of the WPs, these states penetrate deeper into the bulk. 
Consequently, we observe a decrease in the values of $|\psi|^2$, the probability density of the Fermi arc wavefunction, near the surface and a simultaneous redistribution of the weight towards the tail: in Fig.~\ref{fig:loc}, this evolution is shown by the color
ranging from violet (furthest point from the projection) to yellow (closest to the projection). 
Our data are consistent with the divergent penetration length at the WP projections predicted by theory \cite{Haldane2014}. 

Our numerical results on the localization are crucial for an understanding of surface superconductivity: assuming it pertains to the existence of Fermi arcs~\cite{Kuibarov2023}, superconductivity is confined to 3 or 4 atomic layers at the surface. 
Beyond this point, which corresponds to about $20$ to $25\,\mathrm{\AA}$, the contribution of the Fermi arcs to the density of states becomes too small.

\begin{figure}[tb]
    \centering 
    \includegraphics[width=0.49\textwidth]{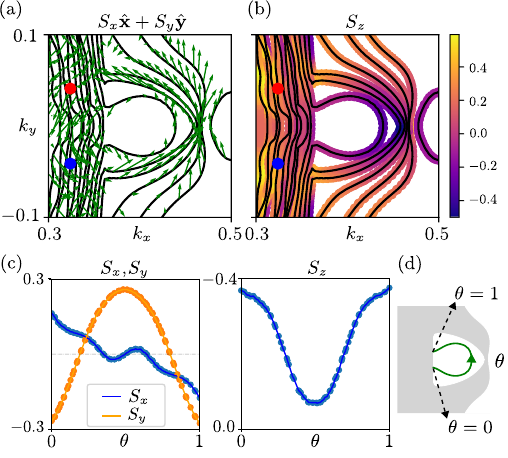}
    \caption{Semi-classical spin structure of the Fermi arcs plotted on top of the Fermi surface of a 20-layer slab. 
    (a) $S_x$ and $S_y$ components represented as arrows, (b) values of $S_z$ are shown with the help of a color-map, and (c), same quantities plotted on a path running along the arc, parameterized by the variable $\theta$, as shown by the sketch in panel (d). 
    Projection of the Weyl points in red (blue).}
    \label{fig:spinarc}
\end{figure}

Finally, we discuss the spin structure of the Fermi arcs. 
In the left panel of Fig.~\ref{fig:spinarc}, we demonstrate how the in-plane component of the spin winds along the arc and aligns with the bulk direction as it approaches the projections of the Weyl points (compare with Fig.~\ref{fig:spin_wp}). 
On the other hand, the out-of-plane spin component $S_z$ points downward, opposite to the orientation of the adjacent bulk states. 
This information is important for scenarios involving scattering by impurities at the surface. 
For instance, in the case of non-magnetic defects and low-momentum processes, no matrix element between the arcs and bulk states will contribute.

In addition, it is worth noting that the spin texture of the Fermi arcs in PtBi$_2$ resembles that of another Weyl semimetal with a layered crystal structure, TaAs~\cite{Xu2016}. 

\section{Effective model}
\label{sec:model}

The extensive degree of hybridization in this compound prevents us from reducing the band structure to a compact Wannier basis, even when our focus is solely on the Weyl points. 
To address more complex topics like superconductivity and, more generally, to satisfy the need for a theoretically simpler model, we now introduce a minimal tight-binding which encodes the topological nature of this system. 
This model consists of two spinful orbitals, and it has been constructed to possess all the crystal symmetries discussed in Sec.~\ref{sec:structure}, while reproducing the same number and distribution of Weyl points found in PtBi$_2$.
We base our approach on the principle that topologically protected Dirac cones can be split into pairs of Weyl points when inversion symmetry is broken. 
To this end, we first establish an appropriate centrosymmetric bulk Hamiltonian, denoted as $H_D$, to which we later add a suitable inversion-breaking term, ensuring the presence of Weyl points.
The  construction of the ``parent'' Hamiltonian $H_D (\mathbf k)$ starts with two uncoupled (spin degenerate) bands, intersecting at circular nodal lines. 
Through the introduction of additional hoppings, these crossing are subsequently reduced to only 6 Dirac cones. 
The desired result is achieved using four adjustable parameters ($\alpha$, $\beta$, $\lambda$, $\mu$). 
Our starting point is
\begin{align}
h_0(&\mathbf k) = [ \mu-\cos k_1-\cos k_2 \nonumber \\
&{}-\cos(k_1+k_2) + \beta \cos k_3 ]\, \Gamma_1 + \beta \sin k_3\, \Gamma_3\nonumber \\
&{}+ \lambda [ \sin k_1  + \sin k_2  - \sin(k_1+k_2) ]\, \Gamma_3.
\end{align}
Following Ref.~\cite{Facio2019}, we have introduced the notation
\begin{gather*}
    \Gamma_1=\tau_z \sigma_0, \quad \Gamma_2 = \tau_x \sigma_x, \quad \Gamma_3 = \tau_y \sigma_0 \\
    \Gamma_{2,j} = \mathcal C_3^j \Gamma_2 \mathcal C_3^{-j} \quad \textrm{with} \quad \mathcal{C}_3 \equiv \tau_0 \exp \left( -i\, \frac {\pi}{3}\, \sigma_z \right).
\end{gather*}
Here, ${\sigma_i}$ are Pauli matrices corresponding to the spin degree of freedom, while ${\tau_i}$ parameterize the orbital degree of freedom. 
The matrix $C_3$ represents rotations along the z-axis by $2\pi/3$. 
Also, we will use the convention $k_i = \mathbf a_i \cdot \mathbf k$, where $\mathbf a_i$ are the lattice vectors $\mathbf a_1 = (0,1,0)$, $\mathbf a_2 = (\sqrt{3}/2,-1/2,0)$, $\mathbf a_3 = (0,0,1)$.

This starting Hamiltonian includes an on-site term $\mu$ and two types of nearest-neighbor hoppings: intra-orbital hoppings (terms multiplied by $\Gamma_1$) and inter-orbital hoppings ($\Gamma_3$). 
The magnitude of the in-plane hoppings between similar orbitals is set to 1, and the hopping between distinct orbitals is given by $\lambda$. 
The model possesses the desired mirror and rotation symmetries discussed in Section \ref{sec:structure}, as well as time-reversal and inversion symmetry, which can be expressed through the matrices
\begin{equation}
    T = i \tau_0 \sigma_y \mathcal K , \quad I = \tau_z\sigma_0,
\end{equation}
where $\mathcal K$ denotes complex conjugation.
The presence of a threefold rotation symmetry gives the condition
\begin{equation}
    \mathcal C_3^{-1} h_0(k_1,k_2,k_3) \mathcal C_3 = h_0(k_2,-k_1-k_2,k_3),
\end{equation}
whereas mirror symmetry along the $k_1=-2k_2$ plane, which is given by the matrix $M_1=i\tau_0\sigma_x$, yields
\begin{equation}
    M_1^{-1} h_0(k_1,k_2,k_3) M_1 = h_0(k_1,-k_1-k_2,k_3).
\end{equation}
The other two mirror symmetries can be obtained by subsequent applications of $\mathcal C_3$ to $M_1$.

\begin{figure}[tb]
    \centering
    \includegraphics[width=0.49\textwidth]{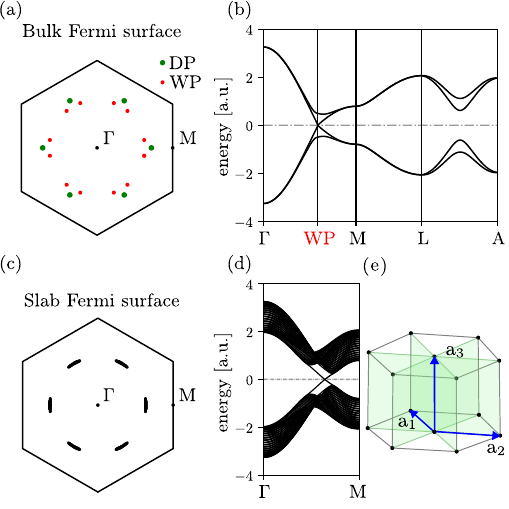}
    \caption{(a) 2D projection of the bulk Fermi surface of our model, comparing $H_D$ (green) and $H$ (red). 
    When inversion is present ($H_D$), 6 Dirac points (DP) are located on the high-symmetry $\Gamma-$M lines. 
    (b) Band structure of the model upon the breaking of inversion symmetry. 
    The path was chosen to show one of the Weyl points. 
    (c) Fermi surface of a finite 30-layers slab stacked along the \textit{z} direction, showing the Fermi arcs. 
    (d) Band structure of the finite slab along the $\Gamma-$M path, again showing the Fermi arcs crossing the Fermi level. 
    (e) Sketch of the geometry of the effective model. 
    All the plots were obtained with the following parameters: $\beta = -0.65$, $\mu=0.4$, $\lambda=1.2$, $\alpha=\gamma=-0.25$.}
    \label{fig:tb}
\end{figure}

For $-(1+\beta) \le \mu \le 3+\beta$ the system is gapless, and for our choice of $\mu$ and $\beta$ (see caption of Fig.~\ref{fig:tb}), when $\lambda = 0$ the band structure possesses two nodal lines (at $k_3=0$ and $k_3=\pi$). 
Allowing for $\lambda \neq 0$, they are split into 12 crossings: 6 Dirac cones in the $k_3=0$ plane and another 6 Dirac cones in the $k_3=\pi$ plane.
However, it is important to observe that all terms introduced until now are spin independent, resulting in $h_0$ being block diagonal in terms of spin. 
To address this, we introduce a spin-orbit-coupling term, $\alpha h_1(\mathbf k)$, which couples up spins and down spins in the two different orbitals. 
Ensuring that all crystal symmetries discussed before are preserved, we find
\begin{align}
h_1(\mathbf k) &= (1-\cos k_3) [ \sin k_1\, \Gamma_2 \nonumber \\ 
&\quad {}+ \sin k_2\, \Gamma_{2,1} - \sin(k_1+k_2)\, \Gamma_{2,2} ] .
\end{align}
The presence of this term also splits the redundant crossings at $k_3 = \pi$, leaving us with only 6 Dirac cones. 
We arrive at the Hamiltonian
\begin{equation}
    H_D(\mathbf k) = h_0 (\mathbf k) + \alpha h_1(\mathbf k).
    \label{eq:model_withdirac}
\end{equation}
Up to now, both time-reversal as well as inversion symmetry are present, excluding the existence of Weyl points, and forcing bands to be doubly degenerate at any momentum. 
Additionally, $H_D$ is also characterized by the presence of chiral symmetry, here represented by the matrix $\Theta = \tau_x \sigma_z$, which anticommutes with the Hamiltonian, $ \{ \Theta,H_D(\mathbf k) \} = 0$. 

We break inversion through the introduction of an additional \textit{k}-independent term $h_2 = \gamma \tau_x \sigma_0$. 
Consequently, the 6 Dirac cones split into a total of 12 Weyl cones. 
Note that this last term also breaks chiral symmetry. 
The final Hamiltonian is
\begin{align}
H(\mathbf k) &= H_D(\mathbf k) + \gamma \tau_x \sigma_0 \nonumber \\
&= h_0(\mathbf k) + \alpha h_1(\mathbf k) + \gamma \tau_x  \sigma_0 .
\end{align}
The evolution of the band structure is shown schematically in Figs.~\ref{fig:tb}(a) and~(b). 
For $\gamma=0$, the system exhibits Dirac cones along the high-symmetry $\Gamma$--$\mathrm{M}$ lines of the BZ (green point in panel a). 
Upon breaking inversion ($\gamma \neq 0$), the cones are split into 6 pairs of Weyl nodes (red dots) which are located above (and below) the $k_z=0$ plane, as in PtBi$_2$. 
On the right, we show the band structure along a path crossing the Weyl point.

The model exhibits clear Fermi arcs at its surfaces, which we also illustrate in Figs.~\ref{fig:tb}(c) and (d).
In panel (c), we show the 2D Fermi surface of the finite slab, where we can identify the Fermi arcs connecting the projection of the Weyl point [compare with panel (a)], while in panel (d), we show the band structure of a finite slab, where we can discern two distinct types of surface states. 
In the picture wee see both of the arcs belonging to the opposite terminations because of the finite size of the system. 
If we were to make a semi-infinite slab, the surface states associated to the opposite termination would disappear.

\section{Superconducting pairing}
\label{sec:pairing}

In this section, we discuss the superconducting pairing state at a single surface of PtBi$_2$. 
The discussion is based on symmetry analysis and two simplifying assumptions: First, since the Fermi arcs are small compared to the size of the surface Brillouin zone we assume that the superconducting state can be understood by considering the pairing at the midpoint of each arc. 
The pairing elsewhere along the arc is then determined by this point, i.e., the superconducting order parameter is stiff along each arc. 
Note that this does not imply that the order parameter is \emph{constant} along the arc. 
Second, we restrict ourselves to superconducting states that do not break crystalline symmetries since there is no experimental indication for such symmetry breaking. 
Formally speaking, we consider superconducting pairing matrices that transform according to the trivial irreducible representation (irrep) $A_1$ of $C_{3v}$. 
At this level, the discussion pertains to both surface terminations.

The absence of inversion symmetry implies that singlet and triplet pairing generically coexist \cite{Frigeri2004}. 
We thus have to represent the superconducting pairing by a $2\times 2$ pairing matrix acting on the spin Hilbert space, namely the off-diagonal block $\Delta$ of a $4\times 4$ Bogoliubov--de Gennes Hamiltonian. 
(Unlike for the effective bulk model in Sec.~\ref{sec:model}, we do not require an orbital degree of freedom here.) 
This matrix can be written as $\Delta(\mathbf{k}) = D(\mathbf{k}) U_T$, where $U_T=i\sigma_y$ is the unitary part of the time-reversal operator $T$ and $D(\mathbf{k})$, unlike $\Delta(\mathbf{k})$, transforms like a matrix under symmetry operations \cite{TiB21}. 
$D$ can be expanded in the identity and Pauli matrices $\sigma_i$ with generally momentum-dependent coefficients. 
The matrices $\sigma_i$ can be organized as irreducible tensor operators of irreps of $C_{3v}$. 
The irreps can be found by checking the transformation of these matrices under the elements of $C_{3v}$:
\begin{align}
\sigma_0: &\quad A_{1+} , \\
\sigma_z: &\quad A_{2-} ,
\label{1.S3.irrep} \\
( \sigma_y, -\sigma_x ): &\quad E_- ,
\label{1.S12.irrep}
\end{align}
where the subscript $\pm$ refers to the sign under time reversal. 
For the components of the two-dimensional irrep $E$ we use the convention that $(k_x,k_y)$ are basis functions in this order.

Since we do not consider variations along each arc, momentum is simply given by the index $\nu \in \mathbb{Z}_6 = \{0,1,2,3,4,5\}$, where the polar angle of the midpoint of each arc is $\nu \pi/3$. 
The momentum-dependent coefficients are thus mappings from the set $\mathbb{Z}_6$ into the complex numbers $\mathbb{C}$. 
These mappings can be understood as six-component complex vectors and therefore form a six-dimensional vector space over $\mathbb{C}$. 
One can easily obtain a basis consisting of basis functions of irreps, starting from the known basis functions in momentum space:
\begin{align}
1: &\quad A_{1+} ,
\label{1.numaps.3a} \\
\left( \cos \frac{\nu\pi}{3}, \sin \frac{\nu\pi}{3} \right): &\quad E_- , \\
\left( \sin \frac{2\nu\pi}{3}, \cos \frac{2\nu\pi}{3} \right): &\quad E_+ ,
\label{1.numaps.3c} \\
\cos \nu\pi = (-1)^\nu: &\quad A_{2-} .
\label{1.numaps.3d}
\end{align}
By multiplying the coefficient functions and the basis matrices, we can obtain all possible superconducting states. 
The multiplication produces $6\times 4 = 24$ linearly independent matrix-valued functions. 
Using standard rules for products of irreps, one can choose them as basis functions belonging to specific irreps. 
However, only products that are even under time reversal (subscript ``$+$'') satisfy fermionic antisymmetry $\Delta^T(-\mathbf{k}) = -\Delta(\mathbf{k})$ \cite{TiB21}, where the superscript $T$ denotes transposition. 
This condition is satisfied by $12$ functions, which characterize possible superconducting states. 
Of these, only three have full ($A_1$) symmetry. 
The corresponding three contributions to $D(\nu)$ read as
\begin{align}
d_0(\nu) &\equiv \sigma_0 , \label{1.d0.3} \\
d_z(\nu) &\equiv (-1)^\nu\, \sigma_z , \\
d_{xy}(\nu) &\equiv \cos\frac{\nu\pi}{3}\: \sigma_y - \sin\frac{\nu\pi}{3}\: \sigma_x ,
\label{1.dxy.3}
\end{align}
and can be characterized as
\textit{s}-wave spin-singlet pairing,
\textit{f}-wave spin-triplet pairing of Ising type, and
\textit{p}-wave spin-triplet pairing of Rashba type,
respectively.
These three pairing states generically coexist since bilinear terms coupling them in a Ginzburg--Landau functional are allowed by symmetry. 
As noted above, singlet and triplet pairing can mix due to the absence of inversion symmetry.

The three states span a three-dimensional space over $\mathbb{C}$ of possible superconducting order parameters of full symmetry. 
Which state in this three-dimensional space is actually realized depends on microscopic details, in particular on the spin structure of the Fermi-arc states.

Figure \ref{fig:spinarc} shows that the spin polarization of the arc $\nu=0$ on the positive $k_x$ axis has a component in the positive \textit{y} direction and a component in the negative \textit{z} direction. 
We parameterize this state by the Bloch vector
\begin{equation}
\frac{\hbar}{2} \begin{pmatrix} 0 \\ \lambda_{xy} \\ -\lambda_z \end{pmatrix} ,
\end{equation}
where $\lambda_{xy}, \lambda_z > 0$ and $\lambda_{xy}^2 + \lambda_z^2 = 1$. 
$\lambda_{xy}$ ($\lambda_z$) describes the relative strength of the Rashba (Ising) spin-orbit coupling. 
The corresponding spin state is
\begin{equation}
|{\uparrow}\rangle_0 = \frac{1}{\sqrt{2}} \begin{pmatrix}
    \sqrt{1-\lambda_z} \\[0.5ex] i\, \sqrt{1+\lambda_z}
  \end{pmatrix} .
\end{equation}
The state of opposite (time-reversed) spin at the same arc is
\begin{equation}
|{\downarrow}\rangle_0 = T\, |{\uparrow}\rangle_0 = U_T\, |{\uparrow}\rangle_0^* .
\end{equation}
This state is not a surface state at the same surface because there is none at the same momentum. 
It thus must be a linear combination of bulk states and possibly a Fermi-arc state at the opposite surface and has vanishing weight at the Fermi energy. 
The states at the other arcs are determined by the point-group symmetries, for example
\begin{align}
|{\uparrow}\rangle_2 &= C_3\, |{\uparrow}\rangle_0 , \\
|{\uparrow}\rangle_3 &= M_1\, |{\uparrow}\rangle_0 ,
\end{align}
where $C_3 = \exp(-i \pi \sigma_z/3)$ implements a rotation by $2\pi/3$ about the \textit{z} axis and $M_1 = \exp(-i \pi \sigma_x/2) = -i\sigma_x$ describes a rotation by $\pi$ about the \textit{x} axis.

There is only one uniquely determined $A_1$ superconducting state (up to an irrelevant global phase) that involves pairing of only electrons from the Fermi arcs, i.e., at the Fermi energy. 
Its pairing matrix is simply
\begin{equation}
\delta_+(\nu) \equiv |{\uparrow}\rangle_\nu \langle{\uparrow}|_\nu .
\label{2.deltapl.2}
\end{equation}
It can be written as
\begin{equation}
\delta_+(\nu) = \frac{1}{2}\, \big[ d_0(\nu) - \lambda_z d_z(\nu) + \lambda_{xy} d_{xy}(\nu) \big]
\label{6.deltaplus.2}
\end{equation}
in terms of the three contributions in Eqs.~\eqref{1.d0.3}--\eqref{1.dxy.3}. 
A second state
\begin{equation}
\delta_-(\nu) \equiv |{\downarrow}\rangle_\nu \langle{\downarrow}|_\nu
  = \frac{1}{2}\, \big[ d_0(\nu) + \lambda_z d_z(\nu) - \lambda_{xy} d_{xy}(\nu) \big]
\end{equation}
can be constructed that corresponds to the pairing of the high-energy states with reversed spin. 
Finally, the state
\begin{align}
\delta_s(\nu) &\equiv \frac{1}{\sqrt{2}\,i}\, \big(
  |{\uparrow}\rangle_\nu \langle{\downarrow}|_\nu - |{\downarrow}\rangle_\nu \langle{\uparrow}|_\nu
  \big) \nonumber \\
&= \frac{1}{\sqrt{2}}\, \big[ \lambda_{xy} d_z(\nu) + \lambda_z d_{xy}(\nu) \big]
\end{align}
describes interband pairing between Fermi-arc surface states and high-energy states. 
These three states span the three-dimensional space of possible superconducting $A_1$ order parameters. 
All three respect time-reversal symmetry since all coefficients are real.

However, only the state $\delta_+(\nu)$ opens a gap at the Fermi energy and thus has a Cooper logarithm in its free energy and a weak-coupling instability. 
Since there is no indication that the pairing interaction in PtBi\textsubscript{2} is strong (i.e., on the order of the Fermi energy) only this state can be realized.

We thus find that pairing of only the spin states at the Fermi energy is possible, without any pairing at high energies. 
The absence of degeneracy of Fermi-arc surface states favors such superconducting states. 
They are enabled by the absence of inversion symmetry.

To conclude this section, we briefly show that treating a local pairing interaction within BCS theory also leads to the state $\delta_+(\nu)$. 
An attractive local interaction of the form
\begin{equation}
H_\mathrm{int} = V \sum_i (c_{i\uparrow}^\dagger c_{i\uparrow}
  + c_{i\downarrow}^\dagger c_{i\downarrow}) = V \sum_i c_i^\dagger c_i ,
\end{equation}
with $V<0$ and the electron spinor
\begin{equation}
c_i \equiv \begin{pmatrix} c_{i\uparrow} \\ c_{i\downarrow} \end{pmatrix} ,
\end{equation}
can be rewritten in the particle-particle channel as~\cite{BoH17, KBT22}
\begin{equation}
H_\mathrm{int} = \frac{V}{2} \sum_i \big( c_i^\dagger U_T c_i^{\dagger T} \big)
  \big( c_i^T U_T^\dagger c_i \big) ,
\end{equation}
up to bilinear terms, which can be absorbed into the non-interacting part of the Hamiltonian. 
The BCS mean-field decoupling of $H_\mathrm{int}$ only permits an order parameter proportional to $-V \langle c_i^\dagger U_T c_i^{\dagger T} \rangle$, which is the order parameter $d_0(\nu)$ given above.

However, so far we have ignored the spin structure of the Fermi-arc states. 
Writing $H_\mathrm{int}$ in momentum space and restricting the interaction to states close to the arcs, we obtain
\begin{equation}
H_\mathrm{int} = \frac{V}{2N} \sum_{\nu\nu'} \sum_{\mathbf{k}\mathbf{k}'}
  \big( c_{\nu\mathbf{k}}^\dagger\, U_T\, c_{\bar\nu,-\mathbf{k}}^{\dagger T} \big)
    \big( c_{\bar\nu',-\mathbf{k}'}^T\, U_T^\dagger\, c_{\nu'\mathbf{k}'} \big) ,
\end{equation}
where $\bar\nu$ denotes the time-reversal partner of arc $\nu$. 
By projecting the interaction onto the arc states $|{\uparrow}\rangle_\nu$, we obtain
\begin{align}
H'_\mathrm{int} &= \frac{V}{2N} \sum_{\nu\nu'} \sum_{\mathbf{k}\mathbf{k}'}
  \big( c_{\nu\mathbf{k}}^\dagger\, \delta_+(\nu)\, U_T\,
    c_{\bar\nu,-\mathbf{k}}^{\dagger T} \big) \nonumber \\
&\qquad{}\times \big( c_{\bar\nu',-\mathbf{k}'}^T\, U_T^\dagger\,
   \delta_+^\dagger(\nu')\, c_{\nu'\mathbf{k}'} \big) ,
\end{align}
where $\delta_+(\nu)$ is given by Eq.~\eqref{2.deltapl.2}. 
Mean-field decoupling then leads exactly to the $A_1$ pairing state that only involves the Fermi-arc states.

\section{Intrinsic $\pi$ junction}
\label{sec:junction}

\begin{figure}[tb]
    \centering
    \includegraphics[width=0.45\textwidth]{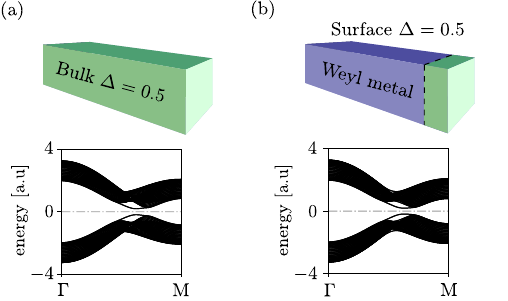}
    \caption{Energy spectrum of the Bogoliubov de--Gennes Hamiltonian of a finite slab with (a) bulk superconductivity and (b) surface-only superconductivity. 
    In both cases, we can see the superconducting gap induced on the Fermi arcs. See text for the details of the calculations.}
    \label{fig:bulksurf_sc_comp}
\end{figure}

In this section, we consider within our effective model the superconducting phase for a general mixture of \textit{s}-wave, \textit{p}-wave, and \textit{f}-wave channels, as discussed in the previous section. 
In the case of bulk superconductivity the Weyl points as well as the Fermi arcs will gap out, while a surface-only pairing will gap out the arcs but leave the bulk metallic, as illustrated in Fig.~\ref{fig:bulksurf_sc_comp}.

Since in PtBi$_2$ superconductivity can develop independently on opposing surfaces, it is interesting to consider the scenario where the superconducting paring $\Delta$ shows a phase difference across opposite terminations, which generates an intrinsic superconductor-metal-superconductor (SMS) Josephson junction. 
To this end, we solve the Bogoliubov-de Gennes (BdG) equation for a finite slab geometry along the \textit{z}-axis:
\begin{equation}
 H_{BdG}(\mathbf k) = \begin{pmatrix} H(\mathbf k) & \Delta(\mathbf k) \\ 
    \Delta^{\dag}(\mathbf k) & - H^{T}(-\mathbf k) \\ 
    \end{pmatrix} ,
\end{equation}
where $H$ is the normal state Hamiltonian of a $N$-layer slab, obtained from our effective model. 
The pairing is a block-diagonal matrix $\Delta(\mathbf k)=\bigoplus_i^N \Delta_i(\mathbf k)$ where $ \Delta_i(\mathbf k)$ is the pairing for each $i$-layer of the slab, and there are no inter-layer components. 
Furthermore,
\begin{equation}
\Delta_i(\mathbf k) = |\Delta| e^{i\phi_i}\, \tau_0\sum_{\alpha}c_{\alpha}d_{\alpha}(\mathbf k) U_T, 
\end{equation}
where $\phi_i$ is the phase (which can change layer by layer) and $d_{\alpha}(\mathbf k)$ with $\alpha = 0$, $z$, and $xy$ corresponding to the three types of pairing: \textit{s}-wave, \textit{f}-wave, and \textit{p}-wave, respectively. 
They are chosen to be consistent with Eq.~\eqref{1.dxy.3}:
\begin{align}
d_0(k_1,k_2) &\equiv \sigma_0 , \\
d_z(k_1,k_2) &\equiv \big[\sin(k_1+2k_2) \nonumber \\
& \quad {}-\sin(2k_1+k_2) \nonumber \\
& \quad {}+\sin(k_1-k_2)\big]\, \sigma_z , \\
d_{xy}(k_1,k_2) &\equiv \sin(k_1+2k_2)\, \sigma_y \nonumber \\
& \quad {}-\sin(2k_1+k_2)\, C_3^{-1}\sigma_yC_3 \nonumber \\
& \quad {}+\sin(k_1-k_2)\, C_3^{-2}\sigma_yC_3^2,
\end{align}
with $C_3=\exp(-\frac {i\pi} 3 \sigma_z)$. 
$c_{\alpha}$ are some coefficients determining the relative weight of the types of pairing. 
We choose the generic values $(3/5,-1/5,1/5)$, where the minus sign was introduced to be consistent with Eq.~\eqref{6.deltaplus.2}. 

\begin{figure*}
    \centering
    \includegraphics[width=0.98\textwidth]{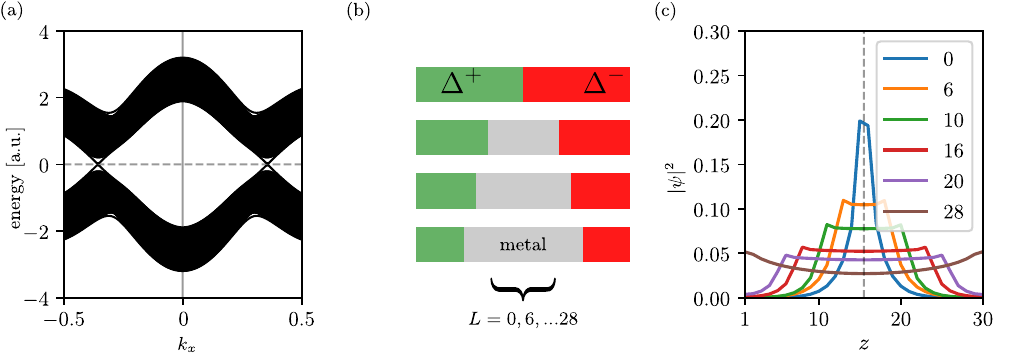}
    \caption{(a) Bogoliubov--de Gennes quasi-particle spectrum for a full $\pi$ junction, corresponding to top case in panel (b). 
    (b) Schematic representation of the junction geometry. 
    Green (red) corresponds to positive (negative) sign of the superconducting pairing $\Delta$. 
    The part of the slab that remains metallic is shown in gray. 
    (c) Probability density of the Andreev bound states as a function of position (layer index) $z$. 
    Different colors refer to the size of the metallic interface, $L$ in panel (b). 
    In all the plots $|\Delta|=0.5$.}
    \label{fig:AndreevLayers}
\end{figure*}

We start by considering the simpler case of a finite slab split in half. 
Each part is a homogeneous superconductor but the pairing differs by a relative phase of $\exp(i\pi)=-1$. 
This phase difference guarantees zero-energy Andreev bound states to appear at the interface between the two superconductors, which we find to be close to the position previously occupied by the Weyl points. 
Since the assumed superconducting order does not break the lattice symmetry, $12$ of such nodes are found. 
The band structure in Fig.~\ref{fig:AndreevLayers}(a) shows two of them.

We can then assume the internal layers to be metallic and check how the Andreev bound states evolve as a function of the thickness of the metallic core, see Fig.~\ref{fig:AndreevLayers}(b). 
For increasing thickness $L$ of the metallic core, the Andreev bound states become progressively less localized, as they spread throughout the metallic phase, with a kink in the wave function noticeable at the superconductor-metal interface. The trend is shown in Fig.~\ref{fig:AndreevLayers}(c), where we plot the probability density $|\psi(z)|^2$ as a function of the \textit{z} coordinate of the slab. 
In Fig.~\ref{fig:AndreevLayers}(b) we sketch the geometries we are considering.
Indeed, we see that the strongest localization happens for the full junction ($L=0$).

\section{Summary and conclusions}
\label{sec:conclusion}

We have used computational methods to characterize the electronic structure of the Weyl semimetal PtBi$_2$, both in the bulk and at the surface, focusing on the Weyl nodes and their topologically protected surface states, the Fermi arcs, investigating their spin texture and penetration depth. 
We also traced the evolution of the Weyl points under externally applied pressure, observing a decrease in energy by $20\,\mathrm{meV}$, as well as on a smooth path between the inversion-broken structure and the energetically close centrosymmetric phase, showing that crystal symmetries enforce the pair merging to happen on the high-symmetry line $\Gamma$--M.

Furthermore, our analysis shows that the complex band structure can be accurately reproduced by a Wannier projection involving spinful Pt $6s$, $5d$, and Bi $6p$ orbitals, resulting in a $72 \times 72$ Hamiltonian matrix.
On this basis, we propose an effective model that captures the distinctive topological features of the system using a minimal set of 4 bands (2 spinful orbitals) while maintaining the appropriate crystalline symmetries, as well as reproducing the correct number and distribution of Weyl points found in PtBi$_2$.
Due to its simplicity, the model may provide a versatile platform for investigating a wider range of phenomena, such as the effects of perturbations or disorder. 

A symmetry analysis of surface superconducting pairings between the Fermi arcs  shows that due to inversion-breaking, coexistence of spin singlet and triplet pairing arise. 
For the superconducting order parameter being stiff along each arc we identify three main channels: homogeneous \textit{s}-wave, \textit{p}-wave, and \textit{f}-wave.  
Finally, we employed our effective model to consider the effect of different phases in the superconducting order parameter on opposite surfaces. 
We showed that, in the limit of a full $\pi$ junction, zero-energy Andreev bound states appear sharply localised at the interface between the two superconducting blocks. 
In the presence of a metallic bulk, the states survive but their wave function rapidly broadens as the size of the metallic core increases. 

With this analysis of the electronic structure of PtBi$_2$ we aimed to set the stage to deepen the understanding of its unique superconducting properties. 
Further theoretical and experimental work is required to establish the microscopic origin, homogeneity, symmetry and mechanism of the surface superconductivity as well as its relation to the topological nature of the Fermi arcs, and the role of bulk inversion symmetry breaking.

\begin{acknowledgments}
We thank Ulrike Nitzsche for technical assistance, Mario Cuoco, Carmine Ortix, and Mattia Trama for fruitful discussions.
We acknowledge financial support by the Deutsche Forschungsgemeinschaft (DFG, German Research Foundation), through SFB 1143 (Project ID 247310070), projects A04 and A05, Project No.\ 465000489, and the W{\"u}rzburg-Dresden Cluster of Excellence on Complexity and Topology in Quantum Matter, ct.qmat (EXC 2147, Project ID 390858490).
\end{acknowledgments}

%

%

\end{document}